\documentclass [11pt,letterpaper,notoc]{JHEP3}

\usepackage{amsmath, bm, graphicx}
\usepackage{epsf}
\usepackage{amscd}
\usepackage{slashed}

\def\ol#1{{\overline{#1}}}

\def\Dslash{D\hskip-0.65em /}

\newcommand{\benn}{\begin{displaymath}}
\newcommand{\eenn}{\end{displaymath}}
\newcommand{\beq}{\begin{equation}}
\newcommand{\eeq}{\end{equation}}
\newcommand{\bea}{\begin{eqnarray}}
\newcommand{\eea}{\end{eqnarray}}
\newcommand{\nn}{\nonumber}

\def\Dslash{{\rlap{\raise 1pt \hbox{$\>/$}}D}}
\def \( {\left(}
\def \) {\right)}

\def\slashchar#1{\ensuremath{                               %
   \setbox0=\hbox{${}#1{}$}       % set a box for #1 
   \dimen0=\wd0                                 % and get its size
   \setbox1=\hbox{/} \dimen1=\wd1               % get size of /
   \ifdim\dimen0>\dimen1                        % #1 is bigger
      \rlap{\hbox to \dimen0{\hfil/\hfil}}      % so center / in box
      {}#1{}                                    % and print #1
   \else                                        % / is bigger
      \rlap{\hbox to \dimen1{\hfil${}#1{}$\hfil}}   % so center #1
      /                                         % and print /
   \fi}}

\title{Sommerfeld enhancement from Goldstone pseudo-scalar exchange }
\author{Paulo F. Bedaque\footnote{bedaque@umd.edu}, Michael I. Buchoff\footnote{mbuchoff@umd.edu} and Rashmish K. Mishra\footnote{rashmish@umd.edu}\\
Maryland Center for Fundamental Physics\\
	Department of Physics, University of Maryland,
	College Park, MD 20742-4111}

\abstract{ We point out that the exchange of a Goldstone pseudo-scalar can provide an enhancement in the dark matter annihilation rate capable of explaining the excess flux seen in high energy cosmic ray data. The mechanism of enhancement involves the coupling of s and d waves through the tensor force that is very strong and, in fact, singular at short distances. The results indicate that large enhancements require some amount of fine tuning. We also discuss the enhancement due to other singular attractive potentials, such as WIMP models with a permanent electric dipole.}

\begin{document}

\section{Introduction}

One of the foremost candidates for dark matter are WIMPS, particles with masses (hundreds of GeV) and interactions in the weak force range. Particles with these properties are generated thermally in the early universe but decouple from the other particles early enough such that a sizable number of them survived at present times. If they are stable their mass and density would have the correct order of magnitude to be the dark matter observed in rotation curves of galaxies, galactic clusters, and cold dark matter models of structure formation. Recently, observations of the cosmic ray flux showed an excess in the high energy component as compared to the standard models \cite{cr_model} describing the diffusion of cosmic rays  originating from supernova remnants \cite{heat,ams,atic,pamela,fermi}. Furthermore, the enhancement is observed in the electron flux but not in the proton flux. One possible explanation to the anomalous cosmic ray data is that it is the result of dark matter particles annihilating \cite{kane,hooper,cholis} into light particles. If these light particles have a mass below hadronic scales, they will decay mainly into electrons which would explain why the enhancement is not seen in the proton flux. One problem with this interpretation of the data is that, given the known dark matter density and the weak scale annihilation cross sections, the electron flux falls short from the observed flux by one to three orders of magnitude (in the absence of significant dark matter clumping and dependent on the observational data set considered). It has been advocated \cite{nima} that models including a long range attractive force between dark matter particles can enhance the annihilation cross sections to the observed levels, a phenomenon refered to as the ``Sommerfeld enhancement"\cite{hisano1,hisano2,cirelli1,cirelli2,cirelli3}. If this force arises from the exchange of a particle with a mass smaller than hadronic scales but larger than the electron mass scale, the standard success of Big-Bang nucleosynthesis is preserved. This light particle (referred here on as $\phi$) can also be the product of dark matter annihilation, nicely explaining the enhancement of the electron but not the proton flux. If this picture is correct, the cosmic ray data should have a peak around the mass of the dark matter particle. This is seen in the older data but not in the most recent GLAST results \cite{fermi}.  An enhancement caused by the exchange of a light boson is also the essential ingredient of a unified theory for dark matter\cite{nima,Fox:2008kb} that also  addresses other observations like the WMAP haze and the DAMA results.

Assuming that the picture above is correct, one has to deal with the fine tuning required for the presence of a particle with mass in the GeV range in a theory with a typical scale in the TeV range. Although solutions to this problem involving supersymmetry exist (for example, see Ref.~\cite{Nomura:2008ru,Mardon:2009gw,Katz:2009qq}), a simple, natural solution would be to assume that the $\phi$ particle is a pseudo-Goldstone boson. The small mass of $\phi$ would be, in this case,  ``technically natural" in the sense that radiative corrections would be zero in the $m_\phi=0$ limit. The force generated through the exchange of a pseudo-scalar pseudo-Goldstone boson is, in general, very complicated and spin dependent. In fact, it is a close analogue to the force between two nucleons generated by pion exchange. Let us assume the dark matter particles are spin $1/2$ Dirac fermions. The force due to $\phi$ exchange, just like the nuclear force due to pions, contains two pieces. One is a central force of the Yukawa type\footnote{For more details on enhancement from a dominant central Yukawa interaction, see Ref.~\cite{Iengo:2009ni,Iengo:2009xf}.} . The other, the tensor force, operates only on spin triplet states and mixes $L=0$ and $L=2$ components of the wave function.  At short distances, $r \ll 1/m_\phi$, the tensor potential goes as $\sim 1/r^3$ and it becomes the dominating piece of the interaction. An observer might naively think that a $1/r$ Yukawa potential will be dominant over a larger range than a $1/r^3$ tensor potential.  However, this is not the case.  When comparing the form of the Yukawa potential to the tensor potential,
\beq
\frac{m_\phi^2 M \alpha }{r}e^{-m_\phi r}\quad \text{vs.} \quad \frac{M \alpha}{r^3}e^{-m_\phi r},\nn
\eeq
one finds that the Yukawa could only dominate when $r>1/m_\phi$.  However, both potentials are exponentially suppressed when $r \sim1/m_\phi$.  Thus, within the range of these potentials, the tensor part dominates the Yukawa force.

An additional complication is that the singular tensor force makes the Schr\"odinger equation ill-defined. Therefore, we are led then to a renormalization procedure to make the non-relativistic effective theory considered here meaningful. In the present case, the renormalization program consists of introducing, in addition to the long range force, a short distance interaction generated by a contact, four-fermion operator in the low-energy effective theory. The range of this interaction is of the order of the inverse mass of other states in the theory that have been integrated out, that is, of the order of $\text{TeV}^{-1}$. The strength of this potential can be determined only through a matching calculation to the full theory in the dark matter sector and, as such, it is an additional parameter from the point of view of the effective theory. The complications due to the mixing of the partial waves and the need for the renormalization procedure form the bulk of the present paper. In the next section, the issues arising in the calculation of the Sommerfeld enhancement in the presence of a singular potential are discussed in the simpler context of a central, single-channel $1/r^3$ potential. This example is not entirely without physical interest as models of dark matter with electric dipole moments would contain such potential. In section 3 we present the results of the full coupled-channel calculation. The conclusions are summarized in section 4.

\section{Sommerfeld enhancement with single-channel singular potential}

The Sommerfeld enhancement is a simple quantum-mechanical effect that, at least in the simpler case of uncoupled channels, has been discussed in the literature several times. Here we briefly review the single channel calculation in order to highlight a couple of points that are relevant to the non-central, singular potential case of interest.

 Let us consider the annihilation of a pair of Dirac particle/antiparticle of mass $M$, attracting each other with a central, spin-independent potential proportional to $1/r^3$. This potential could be generated by the exchange of an electromagnetic or ``dark" photon between particles with a permanent electric dipole.\footnote{Recent studies \cite{Heo:2009xt, Masso:2009mu} found bounds on the electric and magnetic dipole moments from experiments. In the context of our study, these correspond to $(10^{-10} \alpha_e) \lesssim G \lesssim (1 \ \alpha_e)$ where $\alpha_e = 1/137$. Bounds on a dominant dipole interaction in the dark sector from a massive dark photon have yet to be explored.}.      In addition, there will be a short distance interaction at a scale $R$ comparable to the one set by the mass ($R\sim  1/M$). The inclusion of this short distance piece is dictated by the renormalization properties of the singular potential. If we take the quantum mechanical potential model as an effective theory of a microscopic relativistic theory, this short distance potential arises from a four-fermion operator needed to match the high energy theory.
 
\begin{figure}[t]\label{fig:annihilation}
\centering
\includegraphics[width=0.4 \textwidth]{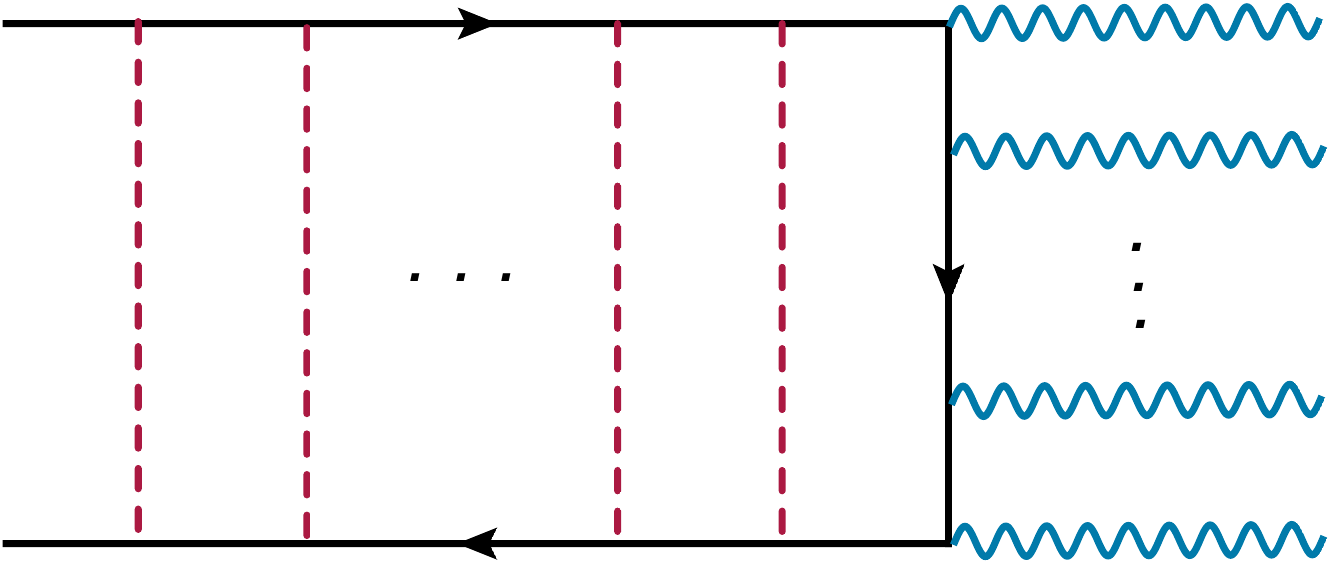}
\caption{Annihilation process enhanced by multiple Goldstone boson exchanges. Solid lines represent the dark matter particles, dashed lines the Goldstone boson and wavy lines may or may not be the Goldstone particle.}
\end{figure}

   We look for a solution to the Schroedinger equation satisfying the usual asymptotic condition at large separations
 \beq\label{eq:asymp}
 \psi(\vec{r}) \rightarrow e^{i k z} + \frac{e^{ikr}}{r}f(\theta)
 =\sum_{l=0}^\infty (2l+1) P_l(\cos\theta) 
 \left[  \frac{e^{ikr}(1+2ik f_l)-(-1)^l e^{-ikr}}{2ikr}\right]
 \eeq where $r$ is the separation between the two particles, $k$ the incoming momentum and $f=\sum_{l=0}^\infty (2l+1) f_l P_l(\cos\theta)$ is the scattering amplitude. The wave function can be expanded in partial waves as
 \beq
 \psi(\vec{r}) = \sum_{l=0}^\infty (2l+1) \frac{u_l(r)}{r} P_l(\cos\theta),
 \eeq and for central, spherically symmetrical potentials, the Schroedinger equation  decouples into separate equations for each partial wave:
 \beq\label{eq:sch_single}
 -u_l''(r) + MV(r) u_l(r) = k^2 u_l(r).
 \eeq The solutions to eq.~(\ref{eq:sch_single}) behave at large $r$ as free waves
 \beq
 u_l(r) \rightarrow A \sin(kr-l\pi/2+\delta_l).
 \eeq The constants $A$ and $\delta_l$ are fixed by the two boundary conditions.
 One is $u_l(0)=0$, needed in order for $\psi(0)$ to be finite. The second one comes from the asymptotic condition in eq.~(\ref{eq:asymp}), which implies 
 \beq\label{eq:single+norm}
 A = \frac{i^l e^{i\delta_l}}{k}.
 \eeq The scattering amplitude is related to the phase shifts $\delta_l$ through $e^{2i\delta_l} = 1+2ik f_l$ and does not depend on the normalization factor $A$. The same is true about the annihilation cross section enhancement. This enhancement is given by the ratio between the interacting and the free wave functions at origin (assuming a zero range bare annihilation amplitude)
 \beq\label{eq:S}
 S = \left| \frac{\psi(0)}{\psi^{(0)}(0)} \right|^2.
 \eeq Thus, in the single channel case, the normalization in eq.~(\ref{eq:single+norm}) plays no role and can be safely taken to be $A=1$ (both for the free  $u^{(0)}$ and interacting wave $u$).  In the expression for the Sommerfeld enhancement above, it was assumed that the annihilation proceeds through a zero range interaction. More realistically, the mechanism of annihilation occurs within a distance scale $\sim 1/M$. This would be true, for instance, if the annihilation goes through the process depicted in Fig.~(\ref{fig:annihilation}) as the vertical internal fermion lines are off-shell by an amount $\sim M$. The annihilation rate would then be determined by a weighted average of the wave function over a region of size $\sim 1/M$ around the origin. However, for well behaved potentials, the wave function varies little over this region and eq.~(\ref{eq:S}) suffices.
 
 %%%%%%%%    FIGURE annihilation graph  %%%%%%%%%%%

%%%%%%%%%%%%%%%%%%%%%%%%%%%%%

 As mentioned above, there is an additional issue when the potential is more singular at origin than $1/r^2$. The 
 Schr\"odinger equation with this potential is meaningless and leads to the ``fall-to-the center" scenario due to the extreme short distance attraction. Bound states exist with arbitrarily high (negative) energy and the hamiltonian is not bounded from below. The way out of this problem is well known. The quantum mechanical model we are considering should be viewed as the low energy effective theory of the underlying relativistic field theory describing the system. The validity of the non-relativistic description and the $1/r^3$ potential is limited to distances larger than a certain scale $R$ where the effective theory breaks down and the potential for shorter distances cannot be trusted. This fact, of course, does not invalidate the use of the Schr\"odinger equation to describe low energy annihilation. It only means that a renormalization procedure should be carried out.  In the present case we carry out the renormalization program by splitting the effective potential into two pieces. The first, valid for $r>R$, is the $1/r^3$ potential generated by a light particle exchange. The second, valid for $r<R$ is a constant potential generated by a four-fermion term in the effective theory lagrangian. The distance $R$ is not a parameter of the model and identical descriptions of the low energy physics should result from different values of $R$, as long as the low energy couplings of the effective theory are made to depend on $R$ appropriately. For non-singular potentials, the dimension 6 four-fermion operator describing the short distance interaction is irrelevant and appear only at higher orders of a low energy expansion. For potentials more singular than $1/r^2$ this is not true and the short distance potential is necessary for a consistent calculation\footnote{Additionally, this short distance potential can have an $\mathcal{O}(k^2)$ correction, but for small $k$, this correction is negligible compared to $V_0$. }. The height $V_0$ and the range $R$ of the potential are arbitrary. However, the requirement that the low energy observables are independent of  the regularization procedure (the values of $R$ and $V_0$) fixes the dependence of $V_0$ on $R$. The whole arbitrariness of the procedure is then reduced to the arbitrariness of one parameter which can be taken to be the value of $V_0$ at one specific value of $R$. Physical results will depend on the value of this parameter. Given a microscopic model one can calculate the scattering amplitudes by matching the microscopic model to our effective theory.  As a consequence, the Sommerfeld enhancement will not have the universal character it has when only non-singular potentials appear. 
 
 The fact that the all low energy observables are rendered cutoff independent by adjusting one short distance coefficient is shown, in the context of singular potential in quantum mechanics, in Ref.~\cite{bedaque_singular,bedaque_towards}. In our case, we solve eq.~(\ref{eq:sch_single}) with the potential
 \beq
 V(r) = \left\{ \begin{matrix}
     -\frac{G}{M^2 r^3}\ \  {\rm for} \ \ r>R,\\
    V_0 \ \  {\rm for} \ \ r<R.
           \end{matrix} \right.
 \eeq  We will consider here a purely s-wave bare annihilation amplitude of short range. ``Short range", in this context, means a range comparable to the cutoff $R$. In principle it is possible to construct models where the range of the short distance elastic interaction, annihilation amplitude, and mass are given by different scales. We will only consider here the more natural case where both ranges are comparable to $R\sim 1/M$. The annihilation amplitude is subjected to the same renormalization procedure described for the elastic interaction. The bare annihilation amplitude is a cutoff dependent quantity and only the amplitude dressed by the potential is physical and cutoff independent. Again, given a microscopic model one can evaluate the value of the annihilation amplitude in that model. For our purposes, we regulate in the annihilation amplitude in a manner similar to the elastic potential
 \beq
 S = \left|  \frac{u(0)}{u_0(0)}  \right|^2 \rightarrow \left|  \frac{\int_R^{3R} dr\ u^2(r)e^{-r^2/R^2}}{\int_R^{3R} dr\ u_0^2(r)e^{-r^2/R^2}} \right|
 \eeq where $u_0(r)$ is the wave function in the absence of the potential with the same normalization as $u(r)$ at large $r$. Different ways of smearing the ratio $u(0)/u_0(0)$ lead to very similar results. The numerical solution of eq.~(\ref{eq:sch_single}) does not bring additional difficulty since the form of the wavefunction is fixed by the $u(0)=0$ boundary condition.
\begin{figure}[t]
\centering
\includegraphics[width=0.5 \textwidth]{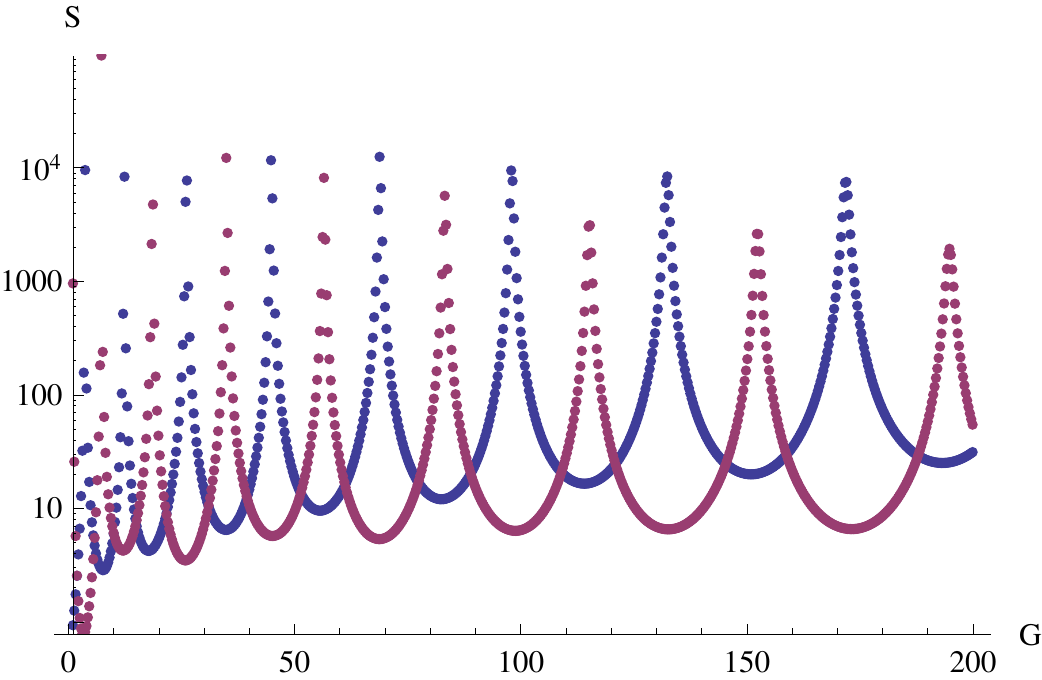}
\caption{Sommerfeld enhancement $S$ as a function of the potential strength $G$ for two value of $V_0=1/MR^2, 10/MR^2$. The other parameter used are $M=1000 \ \text{GeV}, m=1 \ \text{GeV}, k=0.2\ m$ and $R=1/M$.}\label{fig:single}
\end{figure}

%%%%%%%%%%%%%%%%%%%%%%%%%%%%%

  In  Fig.~(\ref{fig:single}) we present results for a sample of parameter values. These parameters  were chosen to mimic the physics of an actual exchange of Goldstone bosons in a theory for dark matter as mentioned in the introduction. Thus, the strength of the potential ($G\sim 1$) is of the order of the one generated by Goldstone bosons with a decay constant of order $\sim M$, and the hierarchy between the masses of the dark matter particle and the Goldstone boson is TeV/GeV  $\sim 1000 $.     A general trend towards larger enhancements for larger values of $G$ is observed. This is expected as a larger value of $G$ means a more attractive potential. In addition there are  wild oscillations superposed on this general trend as either $G$ or $V_0$ is varied. The peaks in the enhancement occur when there is a bound state close to threshold.  Let us compare these results to the enhancements generated by other central potentials. In the case of the Yukawa potential,  analytical estimates are available and show a general trend of $S$ growing proportionally to the strength of the potential. Numerical calculations show that, in addition to this trend, there are regions of parameter space where the enhancement is much larger. The enhancement due to a square well potential is analytically calculable and again shows the large enhancements as bound states cross the threshold. When both long and short distance potentials are combined it is difficult to obtain any reliable analytical estimates. There is also a complicated dependence on the momentum $k$. Since the $1/r^3$ potential does not make sense by itself, we always have to consider the short potential with it, which makes analytical estimates difficult.

\section{Sommerfeld enhancement though Goldstone boson exchange: singular couple-channels}

Goldstone-like pseudo-scalars with derivative couplings to fermions in the dark sector give a naturally light exchange particle for dark matter interactions.    It would be convenient for building dark matter models if such as mechanism could lead to an enhancement of annihilation cross sections.  This possibility was considered  in Ref.~\cite{nima} but quickly discarded since the part of the potential surviving in the $m=0$ limit (the tensor force)  vanishes when averaged over the s-wave initial and final state.  While true that the tensor contribution leads to a change in the orbital angular momentum of $\Delta L = 2$, for a fixed total angular momentum $J$, this contribution can take a state with $L=\ell$ to $L=\ell+2$ and then back to $L=\ell$. Since this process can occur an arbitrary number of times, a coupled-channel analysis is necessary in order to account for this effect as in the one pion exchange potential in the deuteron channel.

The exchange of pseudo-Goldstone bosons between spin-1/2 fermions generates a singular potential similar to the one considered in the previous section. In fact, this problem is very similar to the scattering of two low-energy nucleons and we refer to the literature \cite{bedaque_towards,bedaque1,bedaque2,phillips,epelbaum} for a more detailed discussion of the power counting for this non-relativistic effective theory. In that scenario, the long distance potential is also dominated by the exchange of pseudo-Goldstone particles (pions). There are, however, a few differences between the present problem and the nuclear case. First, the sign of the one-Goldstone exchange potential between particle and anti-particle is the opposite to the one between two particles. Second, the separation between the $M$ and $m$ scales is more pronounced than in the nuclear case.

At the lowest order in the low-momentum expansion the  pseudo-scalar Goldstone boson couples to the axial current of the dark matter particle,
\beq\label{eq:Axial}
\frac{g}{\sqrt{2}} \ol \chi \gamma^\mu  \gamma^5 \chi \frac{\partial_\mu \phi}{f} ,\eeq
where $\chi$ is the dark matter particle, assumed to be a spin 1/2 Dirac fermion, $\phi$ the Goldstone boson,  $g$ the axial coupling of these fermions and $f$ the Goldstone boson decay constant.  Naturality arguments suggest a value of $g\sim1$ and $4\pi f\sim$ TeV, as well as $M \sim $ TeV. The constraints mentioned above and discussed in \cite{nima} imply in $m \gtrsim 0.3$ GeV.
Let us examine the one-Goldstone exchange between a fermion and anti-fermion in our dark sector with total angular momentum $J=1$.   The resulting one-pion exchange potential is well known in nuclear physics and is given by
\beq\label{eq:One_Pi_Pot}
V(r) = V_C(r)\  \sigma_1 \cdot \sigma_2 +V_T(r) \big(3\hat{r}\cdot\sigma_1\:\hat{r}\cdot\sigma_2 - \sigma_1\cdot \sigma_2\big)
\eeq
where
\bea\label{eq:One_Pi_Pot_More}
V_C(r) &=& \alpha m^2 \frac{e^{-m r}}{r} \nn\\
V_T(r) &=& \alpha m^2 \frac{e^{-m r}}{r}\Big(1+\frac{3}{m r} + \frac{3}{m^2 r^2}\Big), \nn
\eea
and $\alpha = g^2/(16\pi f)$.  The central part of the potential corresponds to a $\Delta L = 0$ transition, while the tensor part of the potential corresponds to a $\Delta L = 2$ transition (the spin structure yields a traceless symmetric tensor). In the single channel case, we were interested mainly in the s-channel annihilations since the annihilation rates are larger there due to the absence of a centripetal barrier. Here, we want to consider the same s-channel annihilation. The $J=L=S=0$ channel is of no interest since the Yukawa potential in this state is repulsive.  We are led to consider then the $J=1, S=1, L=0,2$ coupled channels. 
The most general wavefunction with  $S=1, m_J=0$ quantum numbers is \footnote{There is in addition a $J=\ell$ component that, however, does not couple to the $J=\ell-1, \ell+1$ components in the absence of a spin-orbit force.}

\bea
\psi(\mathbf{r}) &=&  \sum_{\ell = 0}^\infty \frac{u_\ell(r)}{r}\Big[ C_{-1}^{(\ell + 1) 0 \ell} Y_{\ell, -1}(\hat{r}) |1,1\rangle + C_{1}^{(\ell + 1) 0 \ell} Y_{\ell, 1}(\hat{r}) |1,-1\rangle + C_{0}^{(\ell + 1) 0 \ell} Y_{\ell, 0}(\hat{r}) |1,0\rangle\Big ]  \nn\\
&+&  \sum_{\ell = 1}^\infty \frac{w_\ell(r)}{r}\Big[ C_{-1}^{(\ell - 1) 0 \ell} Y_{\ell, -1} (\hat{r})|1,1\rangle + C_{1}^{(\ell - 1) 0 \ell} Y_{\ell, 1} (\hat{r})|1,-1\rangle + C_{0}^{(\ell - 1) 0 \ell} Y_{\ell, 0}(\hat{r}) |1,0\rangle \Big ]\nn\\
\eea where the Clebsch-Gordan coefficients, $C^{J m_J L}_{m_L}$, are given by 
\beq
 C_{\pm 1}^{(\ell + 1) 0 \ell} = \sqrt{\frac{\ell}{2(2\ell+1)}}, 
 \quad 
 C_{0}^{(\ell + 1) 0 \ell} = \sqrt{\frac{\ell+1}{2\ell+1}},
 \quad 
  C_{\pm 1}^{(\ell - 1) 0 \ell} = \sqrt{\frac{\ell+1}{2(2\ell+1)}},
  \quad 
  C_{0}^{(\ell - 1) 0 \ell} = -\sqrt{\frac{\ell}{2\ell+1}}.
  \eeq    and the kets represent spin states $|S,m_S\rangle$.
    The asymptotic boundary condition  analogous to Eq.~\eqref{eq:asymp} when the initial particles are in a $m_J=0$ state is 
  \beq\label{double_asymp_bc}
  \psi(\mathbf{r}) \rightarrow e^{ikz} |1,0\rangle + f(\theta,\sigma)\frac{e^{ikr}}{r} =  \frac{1}{2ikr}\sum_{\ell = 0}^\infty (2\ell + 1) P_\ell(\cos \theta) \big[e^{ikr}-(-1)^\ell e^{-ikr}\big]  |1,0\rangle +  f(\hat{r},\sigma)\frac{e^{ikr}}{r}
  \eeq
  where $\sigma$ represents the spin degrees of freedom.   For distances larger than the potential range the wave function is free and approaches
\bea\label{double_asymp_bc_uw}
u_\ell(r) &\rightarrow& B_\ell \sin (kr - l\pi/2 + \epsilon_\ell) = B_\ell \frac{(-i)^\ell}{2i} e^{-i \epsilon_\ell}\Big(e^{ikr+2i \epsilon_\ell} - (-1)^\ell e^{-ikr}\Big) \nn\\
w_\ell(r) &\rightarrow& C_\ell \sin (kr - l\pi/2 + \zeta_\ell) = C_\ell \frac{(-i)^\ell}{2i} e^{-i\zeta_\ell}\Big(e^{ikr+2i\zeta_\ell} - (-1)^\ell e^{-ikr}\Big) 
\eea   
Matching the incoming spherical wave part of Eq.~\eqref{double_asymp_bc} and Eq.~\eqref{double_asymp_bc_uw} we arrive at the relations
%\bea
%B_0 &=& \frac{e^{i\epsilon_0}}{k}\sqrt{4\pi}, \quad \text{no} \; C_0 \nn\\
%0 &=& B_\ell \frac{(-i)^\ell}{2i} e^{i\epsilon_0} C_{\pm 1}^{(\ell + 1) 0 \ell} +C_\ell \frac{(-i)^\ell}{2i} e^{i\epsilon_0} C_{\pm 1}^{(\ell - 1) 0 \ell}\nn\\
%\frac{\sqrt{4\pi(2\ell+1)}}{2ik} &=&  B_\ell \frac{(-i)^\ell}{2i} e^{i\epsilon_0} C_{\pm 1}^{(\ell + 1) 0 \ell} +C_\ell \frac{(-i)^\ell}{2i} e^{i\epsilon_0} C_{\pm 1}^{(\ell - 1) 0 \ell}
%\eea
%The first and third conditions come from matching the asymptotic form in front of $|1,0\rangle$, while the second condition comes from enforcing the $|1,\pm 1\rangle$ state cancels for the incoming spherical wave.  Solving these equations lead to the solution,
\beq\label{B_C_values}
B_\ell =\frac{i^\ell e^{i\epsilon_\ell} }{k} \sqrt{4\pi}\sqrt{\ell + 1} ,\quad
C_\ell = -\frac{i^\ell e^{i\zeta_\ell}}{k} \sqrt{4\pi}\sqrt{\ell} .
\eeq 
Like in the single channel case the  overall normalization of the wave function is unphysical and cancels out in the computation of the Sommerfeld enhancement. The relative normalization of the  $u_\ell$ and  $w_\ell$ components, however, is meaningful. To make things worse, it appears as if the knowledge of the phase shifts $\epsilon_\ell$ and $\xi_\ell$ is needed {\it before} the Scr\"odinger equation can be solved. Fortunately, there is a simple algorithm to construct the wave function with the proper boundary conditions. Let us restrict ourselves to the $J=1,m_J=0$ case of interest. Then only $u_0$ and $w_2$ mix under the influence of the tensor force (from now on we drop the subscript on $u$ and $w$). We first find two linearly independent solutions $(u,w)$ and $(\tilde u,\tilde w)$ satisfying $u(0)=w(0)=0$ and $\tilde u(0)=\tilde w(0)=0$. At large $r$, they will behave like free waves with arbitrary normalizations:
\bea
\begin{matrix}
u(r) &\rightarrow& A_1 e^{-ikr} - B_1 e^{ikr}\nn\\
w(r) &\rightarrow& A_2 e^{-ikr} - B_2 e^{ikr}\nn\\
\end{matrix},
\quad
\begin{matrix}
\tilde u(r) &\rightarrow& \tilde A_1 e^{-ikr} - \tilde B_1 e^{ikr}\nn\\
\tilde w(r) &\rightarrow& \tilde A_2 e^{-ikr} -\tilde B_2 e^{ikr}\nn\\
\end{matrix}.
\eea It is easily verified that the relevant linear combination
\beq\label{eq:linear_comb}
\begin{pmatrix}
u_p(r)\\
w_p(r)
\end{pmatrix}
=
\begin{pmatrix}
u(r) & \tilde u(r) \\
w(r)  & \tilde w(r) 
\end{pmatrix}
\begin{pmatrix}
A_1 & \tilde A_1\\
A_2 & \tilde A_2
\end{pmatrix}^{-1}
\begin{pmatrix}
1\\
-\sqrt{2}
\end{pmatrix}
\eeq satisfies the desired boundary condition eq.~(\ref{double_asymp_bc}). From the solution in eq.~(\ref{eq:linear_comb}) we can calculate the Sommerfeld enhancement. Assuming that the short distance operator describing the annihilation is dominated by the s-wave contribution we calculate
\beq
S = \left|  \frac{\int_R^{3R} dr\ u^2(r)e^{-r^2/R^2}}{\int_R^{3R} dr\ u_0^2(r)e^{-r^2/R^2}} \right|.
\eeq

The resulting Schr\"odinger equation  mixes only states with the same $J$. In the case of the $J=1$ state we have
    
\beq\label{eq:Schro_Pi}
 -\Psi''(r) + M\big( V_L(r)\theta(r-R_0) + V_S(r)\theta(R_0-r)\big)\Psi(r) = k^2  \Psi(r),
 \eeq
 where
 \bea
\Psi = \begin{pmatrix}
u(r)\\
w(r)
\end{pmatrix}, 
V_L(r) =
\begin{pmatrix}
 V_C(r) & \sqrt{8}V_T(r)\\
 \sqrt{8}V_T(r) & V_C(r)-2V_T(r) +6/r^2
 \end{pmatrix},
 V_S(r) =\begin{pmatrix}
- V_0 & 0\\
 0 & 0\end{pmatrix}.
\eea Notice that the short distance potential arises, in the effective theory, from a four-fermion interaction with no derivatives contributing only to the s-wave interaction. Short distance d-wave contributions exist but are of higher order in the low energy expansion.
It is also worthwhile to note that the reversed sign of the potential, as compared to the nuclear case, does not imply that the Sommerfeld enhancement will not occur. In fact, the potential matrix above when diagonalized yields one repulsive and one attractive eigenvalue. Upon inspection, at small $r$, the repulsive eigenvalue is suppressed compared to the attractive one.

\begin{figure}[t]
\centering
\includegraphics[width=0.8 \textwidth]{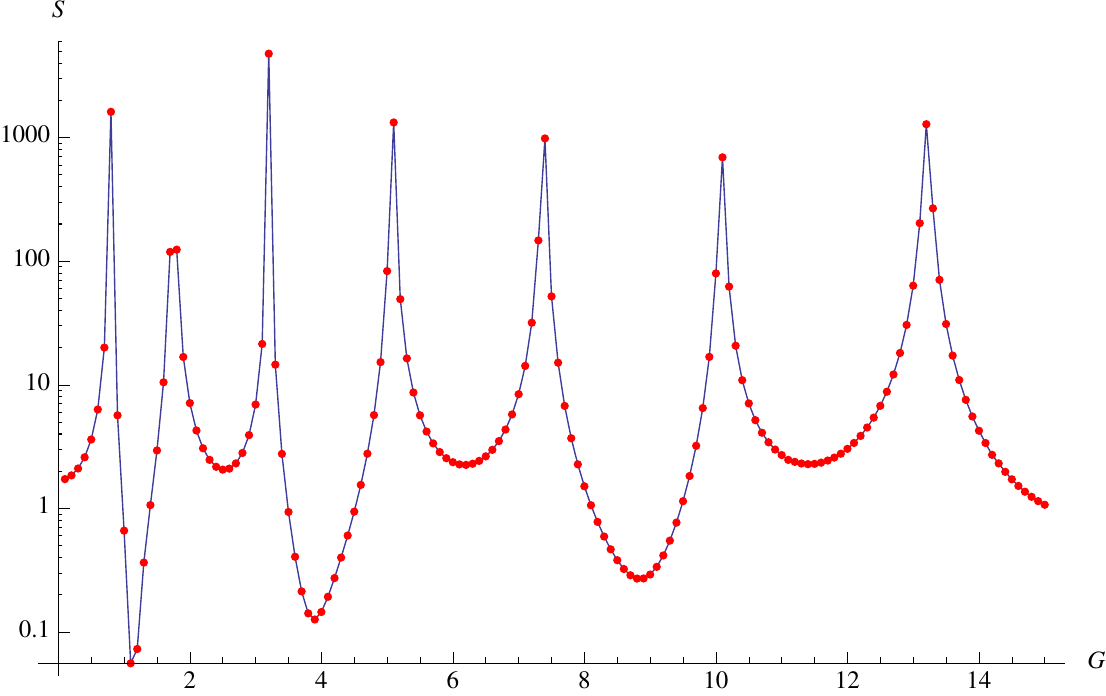}
\caption{Logarithms of the Sommerfeld enhancement $S$ as a function $G$ for $V_0=-10/(M R^2)$, $R=1/M$ and coupling is $\alpha = G/M^2$, where $M=1000 \ \text{GeV}$, and $m=k=1 \ \text{GeV}$. } \label{fig:couple_num_G}
\end{figure}

%%%%%%%%%%%%%%%%%%%%%%%%%%%%%

\begin{figure}[t]
\centering
\includegraphics[width=0.8 \textwidth]{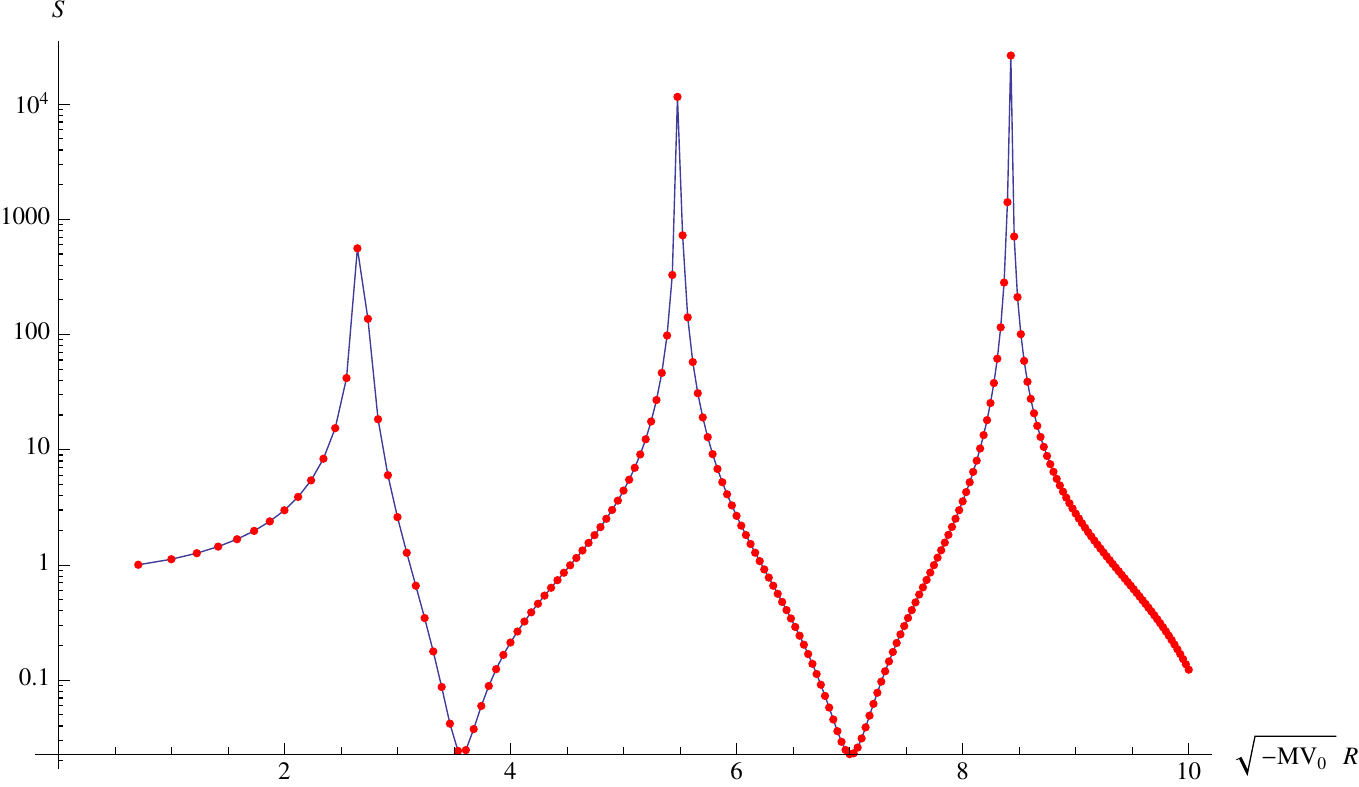}
\caption{ Sommerfeld enhancement $S$ as a function of the short distance potential $V_0$.  For the coupling $\alpha= 1/M^2$, $R=1/M$, where $M=1000 \ \text{GeV}$, and $m=k=1 \ \text{GeV}$.}\label{fig:couple_num_V}
\end{figure}

Examples of numerical results for the Sommerfeld enhancement are shown in Fig.~(\ref{fig:couple_num_G}) as a function of the potential strength $G$ and in Fig.~(\ref{fig:couple_num_V}) as a function of the short distance potential $V_0$. They exhibit a  similar oscillatory pattern as the single channel calculation in Fig.~(\ref{fig:single}).  The enhancement is large only at certain combination of parameters where a bound state close to threshold exists. Contrary to the single channel $1/r^3$ potential, there does not appear to exist any trend towards larger enhancement for stronger potentials (larger $G$).  In this sense the coupled channel problem behaves like a short distance potential (keeping in mind that the oscillatory behavior of the enhancement depends on $G$) .

The periodicity of the enhancement as a function of $\sqrt{-MV_0}R$ shown in Fig.~(\ref{fig:couple_num_V}) is easy to understand. The wave function for $r>R$ depends on  the value of $\psi(R)$ and $\psi'(R)$ only, not on the behavior of $\psi(r)$ at distances smaller than the cutoff $R$. The wave function inside the square well, $u(r) \sim \sin(r\sqrt{-MV_0+k^2})$, exhibits this periodic behavior for small values of $k^2$. As a consequence, the value of $R$ needed to keep the low energy physics fixed varies periodically with $V_0$, a feature well known in non-relativistic effective theories \cite{bedaque_singular}.
  While most $V_0$ values lead to moderate enhancement ($S \sim 1-10$), there exists a significant range of $V_0$ values that can lead to $S \sim 10-1000$ and several near resonance points that can lead to $S>1000$.  Since the quantity $V_0$ is dependent on the higher energy physics that has been integrated out, the magnitude of these enhancements via pseudo-Goldstone exchange is not a universal quantity and can only be determined within the context of a specific microscopic model through the matching of the low energy effective theory to its ultraviolet completion. The outcome of this matching is the value of $V_0$ (given a value of the regulator parameter $R$) that can then be used to determine the enhancement $S$ as explained above.

%%%%%%%%    FIGURE Coupled Numerical G  %%%%%%%%%%%

%%%%%%%%%%%%%%%%%%%%%%%%%%%%%

\section{Discussion}
We discussed the enhancement of the annihilation rates of dark matter induced by the exchange of a pseudo-Goldstone boson. From the point of view of model building for a dark sector, this choice has the advantage of naturally generating the required hierarchy between the WIMP mass and the mass of the particle mediating their interaction.
For the case of spin-1/2 Dirac dark matter that we considered, we find modest enhancement for most of the parameter space.  Still, only moderate fine tuning is necessary to achieve the enhancement of order 10 suggested by the more recent FERMI data.  The enhancements produced are qualitatively similar  to the enhancements generated from a short-distance potential, and hence can be large only with a certain amount of fine tuning . It is a thorny issue to quantify how much fine-tuning is required for a certain enhancement. It is not {\it a priori} clear how to assign probabilities in the effective theory parameter space. For instance, should the probabilities be uniform in $V_0$ or $\sqrt{V_0}$ ?

We have not considered any details of model building of the microscopic theory of dark matter that could realize the Goldstone exchange enhancement we discussed. We point out though that an obvious model with the necessary features would be a scaled up version of QCD in the dark sector. The analogy with nuclear physics are then very strong. WIMP's would be the ``baryons" and the $\phi$ particle would play the role of the ``pions". Assuming the existence of more than one light ``quark", we expect a splitting between different ``baryonic" states to be of the order $\Delta M \approx m^2/\Lambda$ ($\Lambda \sim M \sim 4\pi f$ is the scale of this theory). For $\Lambda \approx 1$  TeV, $m\approx 250 $ MeV , $\Delta M \approx 0.1-1$ MeV range. WIMP excited states with similar splittings have been invoked in the ``Exciting dark Matter" \cite{weiner} scenario also incorporated in the unified description advocated in   \cite{nima}. A similar QCD-like theory was proposed in Ref.~\cite{Alves:2009nf}, but, unlike our model, baryon formation is suppressed leaving the dark mesons as the primary dark matter candidate.  

Another question not investigated here but that deserves attention is the possible enhancement of the annihilation rates through  the formation of bound states over cosmological times previous to their decay into light standard model particles \cite{pospelov,russell}. In the absence of very large or infinite range repulsive forces (the analogue of the Coulomb repulsion in nuclei), the formation of very large ``dark nuclei" seem to be possible. This lumpiness of the distribution of dark matter will have a strong effect on any estimate of their annihilation rates.

\begin{acknowledgments}
We thank Z. Chacko, Tom Cohen, Shmuel Nussinov, and Brian Tiburzi for useful discussions.  Additionally, we would like to thank Aastha Jain for additional computational resources.  
P.~F.~B. and M.~I.~B are supported in part by the 
U.S.~Dept.~of Energy,
Grant No.~DE-FG02-93ER-40762. R.~K.~M. is supported in part by the NSF grant No.~PHY-0801323. 
\end{acknowledgments}


\begin{thebibliography}{99}


 %% model of cr propagation
  %\cite{Strong:2001fu}
\bibitem{cr_model}
  A.~W.~Strong and I.~V.~Moskalenko,
  %``Models for Galactic cosmic-ray propagation,''
  Adv.\ Space Res.\  {\bf 27}, 717 (2001)
  [arXiv:astro-ph/0101068].
  %%CITATION = ASRSD,27,717;%%


%%% CR observation

\bibitem{heat}
  J.~J.~Beatty {\it et al.},
  %``New measurement of the cosmic-ray positron fraction from 5-GeV to
  %15-GeV,''
  Phys.\ Rev.\ Lett.\  {\bf 93} (2004) 241102
  [arXiv:astro-ph/0412230].
  %%CITATION = PRLTA,93,241102;%%
  
  %\cite{Aguilar:2007yf}
\bibitem{ams}
  M.~Aguilar {\it et al.}  [AMS-01 Collaboration],
  %``Cosmic-ray positron fraction measurement from 1-GeV to 30-GeV with
  %AMS-01,''
  Phys.\ Lett.\  B {\bf 646}, 145 (2007)
  [arXiv:astro-ph/0703154].
  %%CITATION = PHLTA,B646,145;%%

%\cite{:2008zzr}
\bibitem{atic}
  J.~Chang {\it et al.},
  %``An Excess Of Cosmic Ray Electrons At Energies Of 300.800 Gev,''
  Nature {\bf 456}, 362 (2008).
  %%CITATION = NATUA,456,362;%%


\bibitem{pamela}
  O.~Adriani {\it et al.}  [PAMELA Collaboration],
  %``An anomalous positron abundance in cosmic rays with energies 1.5.100 GeV,''
  Nature {\bf 458}, 607 (2009)
  [arXiv:0810.4995 [astro-ph]].
  %%CITATION = NATUA,458,607;%%

\bibitem{fermi}
  A.~A.~Abdo {\it et al.}  [The Fermi LAT Collaboration],
  %``Measurement of the Cosmic Ray e+ plus e- spectrum from 20 GeV to 1 TeV with
  %the Fermi Large Area Telescope,''
  Phys.\ Rev.\ Lett.\  {\bf 102}, 181101 (2009)
  [arXiv:0905.0025 [astro-ph.HE]].
  %%CITATION = PRLTA,102,181101;%%
  
 

%%%% excess and dark matter annihilation

%\cite{Kane:2002nm}
\bibitem{kane}
  G.~L.~Kane, L.~T.~Wang and T.~T.~Wang,
  %``Supersymmetry and the cosmic ray positron excess,''
  Phys.\ Lett.\  B {\bf 536}, 263 (2002)
  [arXiv:hep-ph/0202156].
  %%CITATION = PHLTA,B536,263;%%
%\cite{Hooper:2003ad}
\bibitem{hooper}
  D.~Hooper, J.~E.~Taylor and J.~Silk,
  %``Can supersymmetry naturally explain the positron excess?,''
  Phys.\ Rev.\  D {\bf 69}, 103509 (2004)
  [arXiv:hep-ph/0312076].
  %%CITATION = PHRVA,D69,103509;%%

%\cite{Cholis:2008qq}
\bibitem{cholis}
  I.~Cholis, D.~P.~Finkbeiner, L.~Goodenough and N.~Weiner,
  %``The PAMELA Positron Excess from Annihilations into a Light Boson,''
  arXiv:0810.5344 [astro-ph].
  %%CITATION = ARXIV:0810.5344;%%
 
 %\cite{nima}
\bibitem{nima}
  N.~Arkani-Hamed, D.~P.~Finkbeiner, T.~R.~Slatyer and N.~Weiner,
  %``A Theory of Dark Matter,''
  Phys.\ Rev.\  D {\bf 79}, 015014 (2009)
  [arXiv:0810.0713 [hep-ph]].
  %%CITATION = PHRVA,D79,015014;%%
 
  %%% sommerfeld enhancement
  %\cite{Hisano:2003ec}
\bibitem{hisano1}
  J.~Hisano, S.~Matsumoto and M.~M.~Nojiri,
  %``Explosive dark matter annihilation,''
  Phys.\ Rev.\ Lett.\  {\bf 92}, 031303 (2004)
  [arXiv:hep-ph/0307216].
  %%CITATION = PRLTA,92,031303;%%
%\cite{Hisano:2004ds}
\bibitem{hisano2}
  J.~Hisano, S.~Matsumoto, M.~M.~Nojiri and O.~Saito,
  %``Non-perturbative effect on dark matter annihilation and gamma ray
  %signature from galactic center,''
  Phys.\ Rev.\  D {\bf 71}, 063528 (2005)
  [arXiv:hep-ph/0412403].
  %%CITATION = PHRVA,D71,063528;%%
%\cite{Cirelli:2008pk}
\bibitem{cirelli1}
  M.~Cirelli, M.~Kadastik, M.~Raidal and A.~Strumia,
  %``Model-independent implications of the e+, e-, anti-proton cosmic ray
  %spectra on properties of Dark Matter,''
  Nucl.\ Phys.\  B {\bf 813}, 1 (2009)
  [arXiv:0809.2409 [hep-ph]].
  %%CITATION = NUPHA,B813,1;%%
%\cite{Cirelli:2007xd}
\bibitem{cirelli2}
  M.~Cirelli, A.~Strumia and M.~Tamburini,
  %``Cosmology and Astrophysics of Minimal Dark Matter,''
  Nucl.\ Phys.\  B {\bf 787}, 152 (2007)
  [arXiv:0706.4071 [hep-ph]].
  %%CITATION = NUPHA,B787,152;%%
%\cite{Cirelli:2008id}
\bibitem{cirelli3}
  M.~Cirelli, R.~Franceschini and A.~Strumia,
  %``Minimal Dark Matter predictions for galactic positrons, anti-protons,
  %photons,''
  Nucl.\ Phys.\  B {\bf 800}, 204 (2008)
  [arXiv:0802.3378 [hep-ph]].
  %%CITATION = NUPHA,B800,204;%%

%%% unified

%\cite{Fox:2008kb}
\bibitem{Fox:2008kb}
  P.~J.~Fox and E.~Poppitz,
  %``Leptophilic Dark Matter,''
  arXiv:0811.0399 [hep-ph].
  %%CITATION = ARXIV:0811.0399;%%

%\cite{Nomura:2008ru}
\bibitem{Nomura:2008ru}
  Y.~Nomura and J.~Thaler,
  %``Dark Matter through the Axion Portal,''
  arXiv:0810.5397 [hep-ph].
  %%CITATION = ARXIV:0810.5397;%%

%\cite{Mardon:2009gw}
\bibitem{Mardon:2009gw}
  J.~Mardon, Y.~Nomura and J.~Thaler,
  %``Cosmic Signals from the Hidden Sector,''
  arXiv:0905.3749 [hep-ph].
  %%CITATION = ARXIV:0905.3749;%%

%\cite{Katz:2009qq}
\bibitem{Katz:2009qq}
  A.~Katz and R.~Sundrum,
  %``Breaking the Dark Force,''
  JHEP {\bf 0906}, 003 (2009)
  [arXiv:0902.3271 [hep-ph]].
  %%CITATION = JHEPA,0906,003;%%

%\cite{Iengo:2009ni}
\bibitem{Iengo:2009ni}
  R.~Iengo,
  %``Sommerfeld enhancement: general results from field theory diagrams,''
  JHEP {\bf 0905}, 024 (2009)
  [arXiv:0902.0688 [hep-ph]].
  %%CITATION = JHEPA,0905,024;%%
    
%\cite{Iengo:2009xf}
\bibitem{Iengo:2009xf}
  R.~Iengo,
  %``Sommerfeld enhancement for a Yukawa potential,''
  arXiv:0903.0317 [hep-ph].
  %%CITATION = ARXIV:0903.0317;%%


    
%\cite{Heo:2009xt}
\bibitem{Heo:2009xt}
  J.~H.~Heo,
  %``Electric Dipole Moment of Dirac Fermionic Dark Matter,''
  arXiv:0902.2643 [hep-ph].
  %%CITATION = ARXIV:0902.2643;%%

%\cite{Masso:2009mu}
\bibitem{Masso:2009mu}
  E.~Masso, S.~Mohanty and S.~Rao,
  %``Dipolar Dark Matter,''
  arXiv:0906.1979 [hep-ph].
  %%CITATION = ARXIV:0906.1979;%%
  
  %%%%%% review nuclear eft
  
   %\cite{Beane:2001bc}
\bibitem{bedaque_towards}
  S.~R.~Beane, P.~F.~Bedaque, M.~J.~Savage and U.~van Kolck,
  %``Towards a perturbative theory of nuclear forces,''
  Nucl.\ Phys.\  A {\bf 700}, 377 (2002)
  [arXiv:nucl-th/0104030].
  %%CITATION = NUPHA,A700,377;%%



  %\cite{Bedaque:2002mn}
\bibitem{bedaque1}
  P.~F.~Bedaque and U.~van Kolck,
  %``Effective field theory for few-nucleon systems,''
  Ann.\ Rev.\ Nucl.\ Part.\ Sci.\  {\bf 52}, 339 (2002)
  [arXiv:nucl-th/0203055].
  %%CITATION = ARNUA,52,339;%%
  
 %\cite{Beane:2000fx}
\bibitem{bedaque2}
  S.~R.~Beane, P.~F.~Bedaque, W.~C.~Haxton, D.~R.~Phillips and M.~J.~Savage,
  %``From hadrons to nuclei: Crossing the border,''
  arXiv:nucl-th/0008064.
  %%CITATION = NUCL-TH/0008064;%%

%\cite{Phillips:2002da}
\bibitem{phillips}
  D.~R.~Phillips,
  %``Building light nuclei from neutrons, protons, and pions,''
  Czech.\ J.\ Phys.\  {\bf 52}, B49 (2002)
  [arXiv:nucl-th/0203040].
  %%CITATION = CZYPA,52,B49;%%
  
  %\cite{Epelbaum:2008ga}
\bibitem{epelbaum}
  E.~Epelbaum, H.~W.~Hammer and U.~G.~Meissner,
  %``Modern Theory of Nuclear Forces,''
  arXiv:0811.1338 [nucl-th].
  %%CITATION = ARXIV:0811.1338;%%

%\cite{Beane:2000wh}
\bibitem{bedaque_singular}
  S.~R.~Beane, P.~F.~Bedaque, L.~Childress, A.~Kryjevski, J.~McGuire and U.~v.~Kolck,
  %``Singular Potentials and Limit Cycles,''
  Phys.\ Rev.\  A {\bf 64}, 042103 (2001)
  [arXiv:quant-ph/0010073].
  %%CITATION = PHRVA,A64,042103;%%


  %\cite{Finkbeiner:2007kk}
\bibitem{weiner}
  D.~P.~Finkbeiner and N.~Weiner,
  %``Exciting Dark Matter and the INTEGRAL/SPI 511 keV signal,''
  Phys.\ Rev.\  D {\bf 76}, 083519 (2007)
  [arXiv:astro-ph/0702587].
  %%CITATION = PHRVA,D76,083519;%%

%\cite{Alves:2009nf}
\bibitem{Alves:2009nf}
  D.~S.~M.~Alves, S.~R.~Behbahani, P.~Schuster and J.~G.~Wacker,
  %``Composite Inelastic Dark Matter,''
  arXiv:0903.3945 [hep-ph].
  %%CITATION = ARXIV:0903.3945;%%
  
  
  %%% WIMPonium
  %\cite{Pospelov:2008jd}
\bibitem{pospelov}
  M.~Pospelov and A.~Ritz,
  %``Astrophysical Signatures of Secluded Dark Matter,''
  Phys.\ Lett.\  B {\bf 671}, 391 (2009)
  [arXiv:0810.1502 [hep-ph]].
  %%CITATION = PHLTA,B671,391;%%
  

  
  %\cite{MarchRussell:2008tu}
\bibitem{russell}
  J.~D.~March-Russell and S.~M.~West,
  %``WIMPonium and Boost Factors for Indirect Dark Matter Detection,''
  Phys.\ Lett.\  B {\bf 676}, 133 (2009)
  [arXiv:0812.0559 [astro-ph]].
  %%CITATION = PHLTA,B676,133;%%




\end{thebibliography}
\end{document}